\documentclass[12pt]{article}
\usepackage{amsmath,amssymb,epsfig,color}
\usepackage{psfrag}
\usepackage{graphicx}% Include figure files
\usepackage{bm}% bold math

\newcommand{\be}{\begin{equation}}  
\newcommand{\ee}{\end{equation}} 
\def\slash#1{#1\!\!\!/\!\,\,}  
\newcommand{\nl}{\nonumber \\ }

\allowdisplaybreaks

\begin{document}

\begin{titlepage}

\begin{flushright}

EFI 11-23\\
%\today
August 1, 2011
\end{flushright}

\vspace{0.7cm}
\begin{center}
\Large\bf Model independent determination of the axial mass parameter
in quasielastic neutrino-nucleon scattering
\end{center}

\vspace{0.8cm}
\begin{center}
{\sc   Bhubanjyoti Bhattacharya, Richard J. Hill and Gil Paz}\\
\vspace{0.4cm}
{Enrico Fermi Institute and Department of Physics \\
The University of Chicago, Chicago, Illinois, 60637, USA
}
\end{center}
\vspace{1.0cm}
\begin{abstract}
  \vspace{0.2cm}
  \noindent
Quasielastic neutrino-nucleon scattering is a basic signal process for 
neutrino oscillation studies. 
At accelerator energies, the corresponding cross section is subject to 
significant uncertainty due to the poorly constrained axial-vector 
form factor of the nucleon. 
A model-independent description of the axial-vector form factor 
is presented. 
Data from the MiniBooNE experiment for quasielastic 
neutrino scattering on $^{12}$C are analyzed under the assumption 
of a definite nuclear model. 
The value of the axial mass parameter, $m_A=0.85^{+0.22}_{-0.07} \pm {0.09}  \, {\rm GeV}$,
is found to differ significantly from extractions based on traditional form factor models. 
Implications for future neutrino scattering 
and pion electroproduction measurements are discussed. 
\end{abstract}
\vfil

\end{titlepage}

\section{Introduction}

High statistics neutrino experiments are probing the hadronic 
structure of nuclear targets at accelerator energies 
with ever greater precision. 
Extracting the underlying weak-interaction parameters, or
new physics signals, requires similar precision in the theoretical 
description of the strong interactions. 

A basic cross section describes the charged-current quasielastic scattering process
on the neutron, 
\be\label{eq:CCQE}
\nu_\mu + n \to \mu^- + p \,. 
\ee
Recent evidence indicates a tension between measurements of this process 
in neutrino scattering at low~\cite{Gran:2006jn,:2007ru,AguilarArevalo:2010zc,AguilarArevalo:2010cx} 
and high~\cite{Lyubushkin:2008pe} neutrino energies,
and between results from neutrino scattering and results 
inferred from pion electroproduction~\cite{Bernard:2001rs}.  
In particular, with a commonly used dipole ansatz for the axial-vector 
form factor of the nucleon,  
\be\label{eq:dipole}
F_A^{\rm dipole}(q^2) = {F_A(0) \over \left[  1 - {q^2/ (m_A^{\rm dipole})^2}  \right]^2 } \,.
\ee
different experiments have reported values for the so-called axial
mass parameter $m_A^{\rm dipole}$. 
World averages reported by Bernard et al.~\cite{Bernard:2001rs} 
find comparable values obtained from neutrino scattering results prior to 1990, 
$m_A^{\rm dipole} = 1.026 \pm 0.021\,{\rm GeV}$, 
and from pion electroproduction, 
$m_A^{\rm dipole} = (1.069 - 0.055) \pm 0.016\,{\rm GeV}$.%
\footnote{ 
The difference $0.055$ is a correction to the conventional representation 
of the 
pion electroproduction amplitude, as predicted by heavy baryon 
chiral perturbation theory~\cite{Bernard:2001rs}. 
}
The NOMAD collaboration reports~\cite{Lyubushkin:2008pe} $m_A^{\rm dipole}=1.05\pm 0.02 \pm 0.06\, {\rm GeV}$. 
In contrast, MiniBooNE reports~\cite{AguilarArevalo:2010zc} 
$m_A^{\rm dipole}=1.35\pm 0.17$ GeV, and other recent results from the 
K2K SciFi~\cite{Gran:2006jn}, K2K SciBar~\cite{Espinal:2007zz}
and MINOS~\cite{Dorman:2009zz} collaborations similarly 
find central values higher than the above-mentioned world average.    
Quasielastic neutrino-nucleon scattering (\ref{eq:CCQE}) is a basic 
signal process in neutrino oscillation studies. 
It is essential to obtain consistency between experiments utilizing 
different beam energies, and different nuclear targets.

While a number of effects could be causing this tension, we here investigate
perhaps the simplest possibility: that the parameterizations  of the
axial-vector form factor in common use are overly constrained.   
Such a possibility seems natural, considering that the dipole ansatz has 
been found to conflict with electron scattering data for the vector form factors. 
We do not offer new insight on whether other effects, such as nuclear modeling, 
could also be biasing measurements. However, we point out that 
by gaining firm control over the nucleon-level amplitude, such nuclear 
physics effects can be robustly isolated.  

The axial mass parameter as introduced 
in (\ref{eq:dipole}) is not well-defined, since the 
true form factor of the proton does not have a pure dipole behavior.   
Sufficiently precise measurements forced to fit this functional 
form will necessarily find different values 
for $m_A^{\rm dipole}$ resulting from sensitivity to different ranges of $q^2$. 
Let us {\it define} the axial mass parameter in terms of the form factor slope at $q^2=0$:  
$m_A=\left[F_A^\prime(0)/ 2 F_A(0)\right]^{-1/2}$. 
This definition is model-independent, and allows us to sensibly address 
tensions between different measurements.   To avoid confusion, whenever (\ref{eq:dipole}) 
is used we refer to the extracted parameter as $m_A^{\rm dipole}$. 
We will show that the slope at $q^2=0$ is essentially the only relevant shape 
parameter for current data at $Q^2\lesssim 1\,{\rm GeV}^2$, 
and introduce the formalism to systematically account for the impact of other 
poorly constrained shape parameters on the determination of $m_A$. 
A related study of the vector form factors of the nucleon was presented in \cite{Hill:2010yb}. 

The paper is structured as follows.  In Section~\ref{sec:dispersion} 
we discuss the application of analyticity and dispersion relations to 
the axial-vector form factor of the nucleon. 
Section~\ref{sec:numerics} presents results for the extraction of the 
axial-vector form factor slope from MiniBooNE data.  
We illustrate constraints imposed by our analysis on nuclear models, 
by determining the binding energy parameter in the Relativistic Fermi Gas (RFG) model
of Smith and Moniz~\cite{Smith:1972xh}.
Section~\ref{sec:piphoto} gives an illustrative analysis of constraints 
on the axial mass parameter from pion electroproduction data. 
Section~\ref{sec:summary} discusses the implications of our results. 
For completeness, Appendix~\ref{sec:appendix} collects formulas 
for the RFG nuclear model.

\section{Analyticity constraints \label{sec:dispersion}}

This section provides form factor definitions and 
details of the model-independent parameterization based on 
analyticity. 

\subsection{Form factor definitions}

The nucleon matrix element of the Standard Model weak charged current is
\begin{multline}\label{eq:current}
\langle p(p^\prime) | J_W^{+\mu} | n(p) \rangle 
\propto 
\bar{u}^{(p)}(p^\prime) \bigg\{ \gamma^\mu F_1(q^2) + {i\over 2 m_N} \sigma^{\mu\nu} q_\nu F_2(q^2) 
\\
+ \gamma^\mu \gamma_5 F_A(q^2) + {1\over m_N} q^\mu \gamma_5 F_P(q^2) 
\bigg\} u^{(n)}(p) \,,
\end{multline}
where $q^\mu = p^{\prime \mu} - p^\mu$, and 
we have enforced time-reversal invariance and 
neglected isospin-violating effects as discussed in Appendix~\ref{sec:appendix}.  
The vector form factors $F_1(q^2)$ and $F_2(q^2)$ 
can be related via isospin symmetry to 
the electromagnetic form factors measured in electron-nucleon scattering. 
At low energy, the form factors are normalized as $F_1(0)=1$, 
$F_2(0)=\mu_p-\mu_n-1$.   For definiteness we take a common nucleon mass, 
$m_N\equiv (m_p+m_n)/2$.  
Parameter values used in the numerical analysis are listed in Table~\ref{tab:inputs}. 
In applications to quasielastic electron- or muon-neutrino scattering, 
the impact of $F_P$ is suppressed 
by powers of the small lepton-nucleon mass ratio.   For our purposes, 
the pion pole approximation is sufficient,%
\footnote{  
Here and throughout, $m_\pi = 140\,{\rm MeV}$ denotes the pion mass. 
}
\be
F_P(q^2) \approx {2 m_N^2 \over m_\pi^2 -q^2} F_A(q^2) \,.  
\ee

The axial-vector form factor is normalized at $q^2=0$ 
by neutron beta decay (see Table~\ref{tab:inputs}). 
Our main focus is on determining the $q^2$ dependence of $F_A(q^2)$ in 
the physical region of quasielastic neutrino scattering, 
$Q^2=-q^2 \ge 0$.    
As discussed in the Introduction, 
an expansion at $q^2=0$ defines 
an ``axial mass parameter'' $m_A$, via
\be\label{eq:mAdef}
F_A(q^2)= F_A(0) \left[ 1 + {2 \over m_A^2} q^2 + \dots \right] 
\implies 
m_A \equiv \sqrt{ 2 F_A(0)\over F_A^\prime(0) } 
\,.
\ee
Equivalently, we may define an  ``axial radius'' $r_A$, via
\be\label{eq:rAdef}
F_A(q^2)= F_A(0)\left[ 1 + \frac{r_A^2}{6} q^2  + \dots \right]  
\implies 
r_A \equiv \sqrt{6 F_A^\prime(0) \over F_A(0) }  
\,. 
\ee
The factors appearing in (\ref{eq:mAdef}) and (\ref{eq:rAdef}) are 
purely conventional, motivated by the dipole ansatz (\ref{eq:dipole}), and 
by the analogous charge-radius definition for the vector form factors.   
Asymptotically, perturbative QCD predicts~\cite{Lepage:1980fj,Carlson:1985zu} 
a $\sim 1/Q^4$ scaling, up to logarithms, for the axial-vector form factor.   
However, the region $Q^2 \lesssim 1\,{\rm GeV}^2$ is far from asymptotic,
and the functional dependence of 
$F_A(q^2)$ remains poorly constrained at accessible neutrino energies. 

\subsection{Analyticity} 

\begin{figure}[h!]
\begin{center}
\psfrag{a}{$\!\!\!\!\!\!\!-Q^2_{\rm max}$} 
\psfrag{b}{$\!\!\!\!9m_\pi^2$} 
\psfrag{t}{$t$}
\psfrag{z}{$\!z$}
\epsfig{file=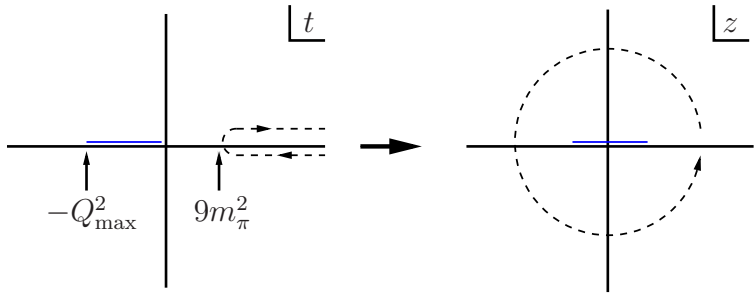,width=10cm}
\caption{\label{cutplane} Conformal mapping of the cut plane to the unit circle.}
\end{center}
\end{figure}  
We proceed along lines similar to the vector form factor analysis 
in \cite{Hill:2010yb}. 
Recall the dispersion relation for the form factor, 
\be\label{eq:dispersion}
F_A(t) = {1\over \pi}\int_{t_{\rm cut}}^\infty dt'\, {{\rm Im}F_A(t' + i0) \over t' - t } \,,
\ee
where $t\equiv q^2$ and the integral starts at the three-pion cut, $t_{\rm cut} = 9 m_\pi^2$.
We can make use of this model-independent knowledge by 
noticing that  the separation between the singular region, $t\ge t_{\rm cut}$, 
and the kinematically allowed physical region, $t\le 0$, implies the existence of 
a small expansion parameter, $|z| < 1$.   
As illustrated in Fig.~\ref{cutplane}, by a standard transformation, we map the
domain of analyticity onto the unit circle in such a way that the
physical region is mapped onto an interval: 
\be\label{eq:zdef}
z(t,t_{\rm cut},t_0) = 
{\sqrt{t_{\rm cut}-t} - \sqrt{t_{\rm cut} - t_0}\over
\sqrt{t_{\rm cut}-t} + \sqrt{t_{\rm cut} - t_0}} \,, 
\ee
where $t_0$ is a free parameter representing the point mapping onto $z=0$. 
Analyticity implies that 
the form factor can be expressed as a power series in the new variable, 
\be\label{eq:zexpansion}
F_A(q^2) = \sum_{k=0}^\infty a_k z(q^2)^k \,. 
\ee
The coefficients $a_k$ are bounded in size, guaranteeing 
convergence of the series. 
Knowledge of ${\rm Im}\,F_A$ over the cut translates into information about the
coefficients in the $z$ expansion \cite{Hill:2010yb}. In particular we have   
\begin{align}\label{eq:fourier} 
a_0&=\frac{1}{\pi}\int_0^\pi d\theta\,{\rm Re}\, F_A[ t(\theta)+i0]=F_A(t_0) \,,
\nonumber\\
a_{k\ge 1}
&=-\frac{2}{\pi}\int_0^\pi d\theta\,  {\rm Im}\, F_A[t(\theta)+i0 ] \,\sin(k\theta) 
= {2\over \pi} \int_{t_{\rm cut}}^\infty {dt\over t-t_0} \sqrt{ t_{\rm cut} - t_0 \over t - t_{\rm cut}} 
{\rm Im}F_A(t)
\sin[ k\theta(t) ] 
\,,
\end{align}
where 
\be
t=t_0+\frac{2(t_{\rm cut}-t_0)}{1-\cos \theta}\equiv t(\theta)\,. 
\ee

\subsection{Coefficient bounds} 

For a given kinematic range $0 \le -t \le Q_{\rm max}^2$, 
we can choose the free parameter $t_0$ in (\ref{eq:zdef}) 
to minimize the resulting maximum size of $|z|$.   
It is straightforward to see that the ``optimal'' value of $t_0$ is 
$t_0^{\rm opt} = t_{\rm cut} \left( 1 - \sqrt{1+Q_{\rm max}^2/t_{\rm cut}} \right)$,
and for this value of $t_0$, 
$|z| \le 
[(1+Q_{\rm max}^2/t_{\rm cut})^{1/4} - 1]/[ (1+Q_{\rm max}^2/t_{\rm cut})^{1/4} + 1 ]$. 
For example, if the kinematic range is $Q_{\rm max}^2 \lesssim 1\,{\rm GeV}^2$, 
then our expansion parameter is constrained to be $|z| \lesssim 0.2$.  
Terms beyond linear order in the expansion are suppressed by $|z|^2 \lesssim 0.04$, etc., 
and are not tightly constrained by current experimental data. 
This is the sense in which the slope of the form factor (conventionally taken at $q^2=0$) 
is essentially the only relevant shape parameter. 
The effects of the higher order terms 
must of course be accounted for in assessing the uncertainty on extracted observables. 
We now turn to this question. 

\begin{table}
\begin{center}
\begin{tabular}{c|cc}
 & $t_0=0$ & \quad $t_0=t_0^{\rm opt}(1.0\,{\rm GeV}^2)$ 
\\
\hline 
$||F_A||_2/|F_A(t_0)|$ & 1.5-1.7 &   1.9-2.3
\\
$||F_A||_\infty/|F_A(t_0)|$ 
& 1.0-1.4 & 1.4-1.8 
\end{tabular} 
\end{center}
\caption{\label{tab:norm}
Typical bounds on the coefficient ratios $ \sqrt{ \sum_k {a_k^2/a_0^2} }$ 
(first line of table) 
and $|a_k/a_0|$ (second line) in an axial-vector dominance ansatz.  
The range 
corresponds to the 
range $250-600\,{\rm MeV}$ for the $a_1$ width and the range $1190-1270\,{\rm MeV}$ for 
the $a_1$ mass. 
} 
\end{table}

The expansion coefficients appearing in (\ref{eq:zexpansion}) 
can be used to define norms,  
\be
|| F_A ||_{p} = \left( \sum_k |a_k|^p \right)^{1/p} \,.
\ee
In particular, 
$|| F_A ||_{\infty} = {\rm sup}_k |a_k| = \lim_{p\to\infty} ||F_A||_p$
provides a bound on the maximum coefficient size. 
The finiteness of the integral appearing in the relation 
\be
||F_A||_2 = \left({1\over\pi}\int_{t_{\rm cut}}^\infty {dt\over t-t_0} 
\sqrt{t_{\rm cut} - t_0 \over t-t_{\rm cut}} |F_A(t)|^2 \right)^{1/2} \,,
\ee
together with $||F_A||_\infty \le ||F_A||_2$, establishes that a finite upper bound exists for 
the coefficients.   
As a first approach to estimating the actual bound $||F_A||_\infty$, 
consider an ``axial-vector dominance'' ansatz, 
$F_A \sim  m_{a_1}^2/ ( m_{a_1}^2 - t - i \Gamma_{a_1} m_{a_1} )$,
where $m_{a_1}=1230(40)\,{\rm MeV}$ and $\Gamma_{a_1}=250-600\,{\rm MeV}$ 
are the mass and width of the lowest lying axial-vector, 
iso-vector meson~\cite{Nakamura:2010}. 
More precisely, let us define the form factor via its dispersion relation 
with~\cite{Schwinger}
\be\label{eq:ImFA}
{\rm Im}F_A(t+i0) ={ {\cal N} m_{a_1}^3 \Gamma_{a_1}  \over (t-m_{a_1}^2)^2  + \Gamma_{a_1}^2 m_{a_1}^2 } \theta(t-t_{\rm cut})\,,
\ee
where ${\cal N}$ is a normalization constant determined below. 
Using the dispersion relation (\ref{eq:dispersion}) with (\ref{eq:ImFA})
we find, 
\be
F_A(t+i0)=\frac{{\cal N}}\pi
\frac{m_{a_1}^3 \Gamma_{a_1}}
{|b(t)|^2}
\Bigg[\frac{1}{2}
\log\left(
\frac{|b(t_{\rm cut})|^2}{|t_{\rm cut}-t|^2}
\right)
+ { m_{a_1}^2 - t \over m_{a_1}\Gamma_{a_1}} \arg[b(t_{\rm cut})] 
+ i\pi \theta(t-t_{\rm cut})\Bigg]\,,
\ee
where $b(t)=t-m_{a_1}^2+ i \Gamma_{a_1} m_{a_1}$, and  
${\cal N}$ is determined by the value of $F_A(0)$.
Table~\ref{tab:norm} displays the values for $||F_A||_2$ and 
$||F_A||_\infty$ computed in this ansatz. 
For the latter quantity one can show that
\be
\left|{a_k \over a_0}\right| 
\le 
\frac{ 2 |{\cal N}|}{|F_A(t_0)|}  
{\rm Im} \left( { - m_{a_1}^2 \over b(t_{\rm cut}) + \sqrt{(t_{\rm cut}-t_0)b(t_{\rm cut})} } 
\right) \,.
\ee
While this model is not a rigorous description of the true spectral function
in (\ref{eq:dispersion}), it indicates an order unity bound on the coefficients
appearing in (\ref{eq:zexpansion}).  
Additional support for an order unity bound is provided by 
a related detailed study of nucleon vector form factors~\cite{Hill:2010yb}, 
and by form factor studies in a wide range of meson transitions~\cite{Hill:2006ub,Bourrely:1980gp}.

In the following numerical analysis, we follow \cite{Hill:2010yb}, 
and investigate fits with various bounds on coefficients, e.g. 
$|a_k|\le 5$ and $|a_k|\le 10$.   

\section{Extraction of the axial mass parameter \label{sec:numerics}}

The MiniBooNE collaboration has presented binned results representing 
the double differential cross section, $d\sigma/dE_\mu d\cos\theta_\mu$, 
for the quasielastic scattering process (\ref{eq:CCQE}) on a neutron bound inside $^{12}$C. 
We apply our description of $F_A(q^2)$ to extract $m_A$ 
(equivalently, $r_A$) from the neutrino scattering data, under the assumption of 
a definite nuclear model, the Relativistic Fermi Gas model~\cite{Smith:1972xh}
as described in Appendix~\ref{sec:appendix}.   

\begin{table}
\begin{center}
\begin{tabular}{l|c|c}
Parameter & Value &  Reference \\
\hline
$|V_{ud}|$ & 0.9742 &  \cite{Nakamura:2010}
\\
$\mu_p$ & 2.793 & \cite{Nakamura:2010}
\\
$\mu_n$ & $-1.913$ & \cite{Nakamura:2010}
\\
$m_\mu$ & 0.1057 GeV & \cite{Nakamura:2010}
\\
$G_F$ & $1.166 \times 10^{-5}$ GeV$^{-2}$ & \cite{Nakamura:2010}
\\
$m_N$ & 0.9389 GeV & \cite{Nakamura:2010}
\\
$F_A(0)$ & $-1.269$ & \cite{Nakamura:2010}
\\
$\epsilon_b$ & 0.025 GeV & \cite{Moniz:1971mt}
\\
$p_F$ & 0.220 GeV & \cite{AguilarArevalo:2010zc}

\end{tabular} 
\end{center}
\caption{\label{tab:inputs}
Numerical values for input parameters. 
} 
\end{table}

Our theory prediction is obtained using
(\ref{eq:doublediff}), integrating over the energy-dependent 
$\nu_\mu$ flux from Table~V of \cite{AguilarArevalo:2010zc}; this result 
is divided by $6$ to obtain the per-neutron event rate, 
and divided by the total flux to obtain the flux-averaged cross section. 
Corresponding experimental
values for the double differential cross section are
taken from Table~VI of \cite{AguilarArevalo:2010zc}.  We form an error
matrix,  \be  E_{ij} = (\delta \sigma_i)^2 \delta_{ij} + (\delta N)^2
\sigma_i \sigma_j  \,, \ee where $\sigma_i = (d\sigma/dE_\mu
d\cos\theta_\mu)\Delta E_\mu \Delta \cos\theta_\mu$  denotes a partial
cross section, $\delta \sigma_i$ denotes the shape uncertainty  from
Table~VII of \cite{AguilarArevalo:2010zc}, and $\delta N = 0.107$ is
the normalization error from \cite{AguilarArevalo:2010zc}.    We form
the chi-squared function  
\be  
\chi^2 = \sum_{ij} (\sigma^{\rm
expt.}_i-\sigma^{\rm theory}_i) E^{-1}_{ij}  (\sigma^{\rm
expt.}_j-\sigma^{\rm theory}_j) \,, 
\ee 
and minimize $\chi^2$ to find best fit values for $m_A$.  
Error intervals are defined by $\Delta \chi^2 = 1$.    
The nucleon form factors and the nuclear
model employ parameter values listed in Table~\ref{tab:inputs}.
Following the analysis of \cite{AguilarArevalo:2010zc},  the vector
form factors $F_1$ and $F_2$ are given by the BBA2003
parameterization~\cite{Budd:2003wb}.  
We use a default value $\epsilon_b=0.025$ GeV, as extracted from electron
scattering data on nuclei in \cite{Moniz:1971mt}.    
This value is different from the central value adopted 
in the MiniBooNE analysis \cite{AguilarArevalo:2010zc}, where
$\epsilon_b=0.034\pm0.09$ GeV. 
We show below that such a high value of $\epsilon_b$ is not favored by the MiniBooNE data, 
but investigate fit results for different values of $\epsilon_b$.  

The slope at $q^2=0$, and hence $m_A$ from (\ref{eq:mAdef}) is most sensitive to low-$Q^2$ data.  
We analyze this sensitivity by considering the effect of a cut on $Q^2$. 
The value of $Q^2$ for a given value of the observed muon energy and angle can be reconstructed 
assuming quasielastic scattering on a free neutron, 
but is not determined unambiguously once nuclear effects are included. 
As a proxy for $Q^2$, we define an approximate ``reconstructed'' $Q^2$,  
\be\label{eq:Q2rec}
Q^2_{\rm rec}=2E^{\rm rec}_\nu E_\mu-2E^{\rm rec}_\nu\sqrt{E_\mu^2-m_\mu^2}\cos\theta_\mu -m_\mu^2 \,, 
\ee 
where $E_\nu^{\rm rec}$ approximates the neutrino energy in the nucleon rest frame, 
\be
E^{\rm rec}_\nu= \frac{m_N E_\mu-m_\mu^2/2}{m_N-E_\mu+\sqrt{E_\mu^2-m_\mu^2}\cos\theta_\mu } \,. 
\ee
We note that $Q^2_{\rm rec}$ coincides with $Q^2_{\rm rec}$ used by K2K in the limit $\epsilon_b\to0$
\cite{Gran:2006jn}, and with $Q^2_{\rm QE}$ used by MiniBooNE in the
limit $\epsilon_b\to0$ and equal proton and neutron masses
\cite{AguilarArevalo:2010zc}. 
For simplicity we have chosen to make the cut independent of the 
binding energy used in the nuclear model. 
We emphasize that this choice is used simply to define the subset of data to be analyzed, and 
does introduce theoretical uncertainty in the numerical results. 

\begin{figure}[h!]
\begin{center}
\psfrag{x}{$Q^2_{\rm max}\,({\rm GeV}^2)$}
\psfrag{y}{\rotatebox{270}{\hspace{-15mm}$m_A(\rm GeV)$}} 
\epsfig{file=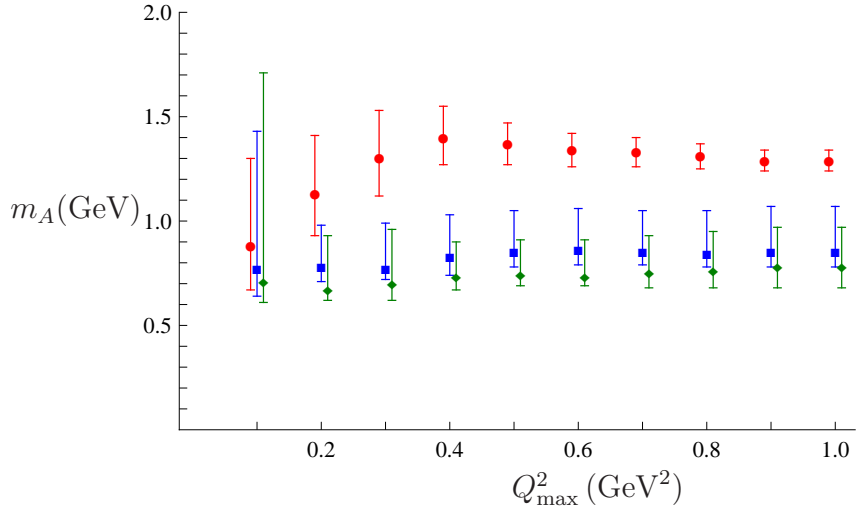,width=10cm}
\caption{\label{fig:mA_Q2} Extracted value of $m_A$ versus $Q^2_{\rm max}$. 
Dipole model results for $m_A^{\rm dipole}$ 
are shown by the red circles;  
$z$ expansion results with $|a_k|\le 5$ are shown by the blue squares, 
$z$ expansion results with $|a_k|\le 10$ are shown by the green diamonds.
}
\end{center}
\end{figure}  

Our results are displayed in Fig.~\ref{fig:mA_Q2}, where 
we compare extractions of $m_A^{\rm dipole}$ in the dipole ansatz (\ref{eq:dipole}) 
with extractions of $m_A$ employing the $z$ expansion (\ref{eq:zexpansion}).  
We present results for data with 
$Q^2_{\rm rec} \le Q^2_{\rm max}$, where $Q^2_{\rm rec}$ is defined in (\ref{eq:Q2rec}) and 
$Q^2_{\rm max}=0.1,0.2,\dots,1.0\,{\rm GeV}^2$.
We study two different coefficient bounds, $|a_k|\le 5$ and $|a_k|\le 10$.   
For definiteness we have truncated the sum in (\ref{eq:zexpansion}) at 
$k_{\rm max}=7$, but have checked that the results do not change significantly 
if higher orders are included. 
As Fig.~\ref{fig:mA_Q2} illustrates, the $z$ expansion results lie systematically below
results assuming the dipole ansatz. 
In contrast to results from the one-parameter dipole ansatz, 
high-$Q^2$ data have relatively small impact on the model-independent determination of $m_A$.  
Taking for definiteness $Q^2_{\rm max}=1.0\,{\rm GeV}^2$, we find
\be\label{eq:MAz}
m_A = 0.85^{+0.22}_{-0.07} \pm {0.09}  \, {\rm GeV}  \qquad \mbox{(neutrino scattering)},
\ee
where the first error is experimental, using the fit with $|a_k|\le 5$, and 
the second error represents residual form factor shape uncertainty, 
taken as the maximum change of the $1\sigma$ interval when the bound is 
increased to $|a_k|\le 10$. 
As a comparison, a fit assuming the dipole form factor, and the same $Q^2_{\rm max}$ 
yields 
$m_A^{\rm dipole} = 1.29 \pm 0.05$~GeV.%
\footnote{
A dipole fit including the entire dataset without a cut on $Q^2_{\rm rec}$ 
yields $m_A^{\rm dipole}=1.28^{+0.03}_{-0.04}$. 
}

It is not our purpose in this paper to investigate in detail 
the additional uncertainty that should be assigned to (\ref{eq:MAz}) due to nuclear effects.   
We note that a fit of the MiniBooNE data to the RFG model 
with free parameter $\epsilon_b$
yields the value, without an assumption on the value of $m_A$, 
(for $Q^2_{\rm max}=1.0\,{\rm GeV}^2$, $k_{\rm max}=7$) 
\be\label{epsilonb}
\epsilon_b = 28 \pm 3 \,{\rm MeV} \,,
\ee
where the result is insensitive to the choice of bound, $|a_k|\le 5$ or $|a_k|\le 10$.%
\footnote{ 
Using a dipole ansatz for $Q^2_{\rm max}=1.0\,{\rm GeV}^2$ 
without fixing $m_A^{\rm dipole}$ yields $\epsilon_b = 22 \pm 7 \,{\rm MeV}$. 
}
While the data do not appear to favor significantly higher values of $\epsilon_b$, 
we note that for $\epsilon_b =34\,{\rm MeV}$~\cite{AguilarArevalo:2010zc}, 
the result (\ref{eq:MAz}) becomes $m_A(\epsilon_b=34\,{\rm MeV})=1.05^{+0.45}_{-0.18}\pm 0.12$, 
compared to $m_A^{\rm dipole}(\epsilon_b=34\,{\rm MeV})=1.44 \pm 0.05$.  

We have performed fits at different values of the parameter $t_0$, finding
no significant deviation in the results. 
The results do not depend strongly on the precise value of the bound 
(e.g. $|a_k|\le 5$ versus $|a_k|\le 10$).   
Similar to \cite{Hill:2010yb}, we conclude that 
the estimation of shape uncertainty in (\ref{eq:MAz}) should be conservative.  
The fit (\ref{eq:MAz}) yields coefficients%
\footnote{
For this purpose we take $k_{\rm max}=7$ in (\ref{eq:zexpansion}) and enforce
$|a_k|\le 10$ for $k\ge 3$.  
} 
$a_0 \equiv F_A(0) = -1.269$, 
$a_1 = 2.9^{+1.1}_{-1.0}$, 
$a_2 = -8^{+6}_{-3}$. 
These values are in accordance with our assumption of order-unity coefficient bounds.  
As discussed in the Introduction, current experiments do not significantly 
constrain shape parameters beyond the linear term, $a_1$.    

\begin{figure}[h!]
\begin{center}
\psfrag{X}{$Q^2$} 
\psfrag{Y}{\rotatebox{270}{\hspace{-20mm}$-F_A(-Q^2)$}}
\epsfig{file=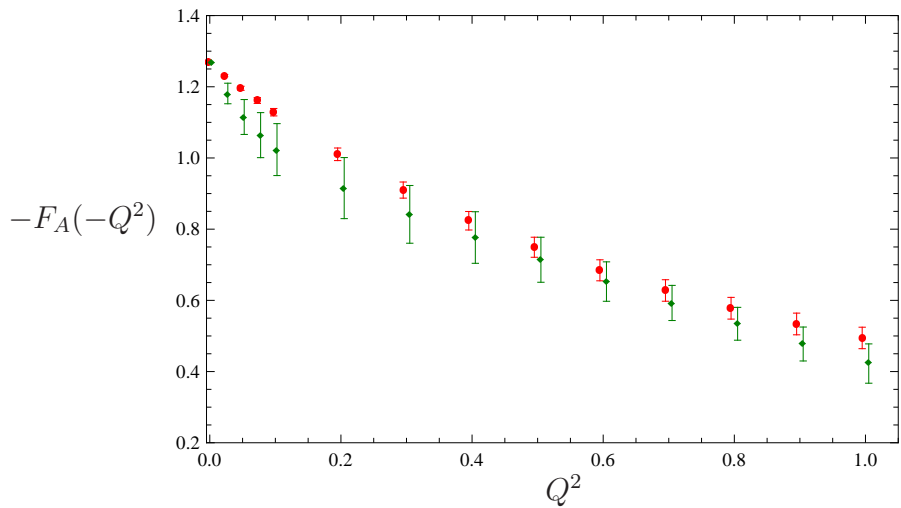,width=10cm}
\caption{\label{fig:ff} Comparison of the axial-vector form factor $F_A$ as
extracted using the $z$ expansion (green diamonds) 
and dipole ansatz (red circles).}
\end{center}
\end{figure}

Figure~\ref{fig:ff} compares the form factor extraction resulting from  
the $z$ expansion fit to the extraction from the 
dipole fit.   Here we take $Q^2_{\rm max}=1.0\,{\rm GeV^2}$, 
$k_{\rm max}=7$ and $|a_k|\le 10$ for the $z$ fit.  
The dipole fit assumes $m_A^{\rm dipole}=1.29\pm 0.05\,{\rm GeV}$.

\section{Comparison to charged pion electroproduction \label{sec:piphoto}} 

\begin{figure}[h!]
\begin{center}
\psfrag{a}{\hspace{-3.5mm}\tiny Average}
\psfrag{b}{\hspace{-3.5mm}\tiny Frascati (1972)~\cite{Amaldi:1972vf}}
\psfrag{c}{\hspace{-3.5mm}\tiny DESY (1973)~\cite{Brauel:1973cw}}
\psfrag{d}{\hspace{-3.5mm}\tiny Daresbury (1975/1976)~\cite{Del Guerra:1975wk,DelGuerra:1976uj}}
\psfrag{e}{\hspace{-3.5mm}\tiny Kharkov (1978)~\cite{Esaulov:1978ed}}
\psfrag{x}{\hspace{-5mm}$m_A(\rm GeV)$}
\epsfig{file=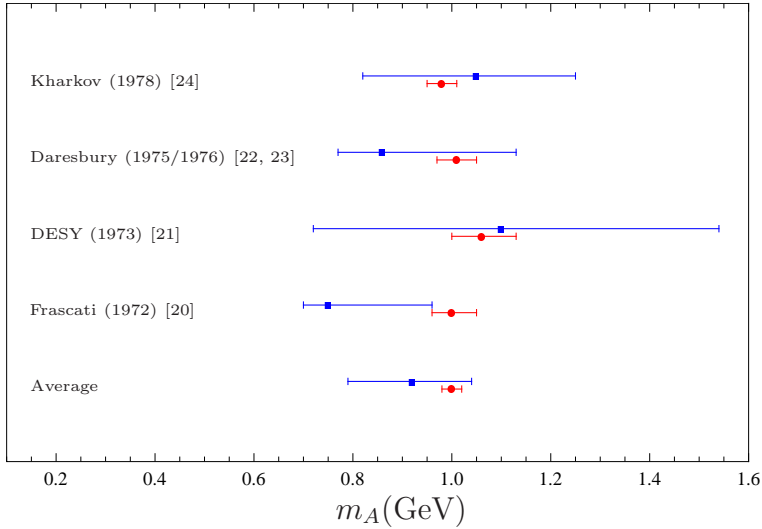,width=10cm}
\caption{\label{fig:piphoto}  Extraction of $m_A$ using charged pion 
electroproduction measurements, in the dipole ansatz and in the $z$ expansion. 
Datasets are as described in the text.   Dipole results are shown as the red circles, 
and $z$ expansion results with $|a_k|\le 5$ are shown as the blue squares.   
}
\end{center}
\end{figure}  

The axial-vector component of the weak current defining $F_A(q^2)$ 
in (\ref{eq:current}) can also be probed 
in pion electroproduction measurements.
The electric dipole amplitude for threshold charged-pion electroproduction 
obeys a low-energy theorem in the chiral limit relating this
amplitude to the axial-vector form factor of the nucleon~\cite{Nambu:1997wa}.  
After applying chiral corrections, such measurements can thus
in principle be used to determine $m_A$.   
Data for this process have been interpreted in the context 
of the dipole ansatz (\ref{eq:dipole}). 
We found that the dipole assumption can strongly bias extractions 
of $m_A$ in neutrino scattering measurements.  
In order to gauge whether the same statement is true for the electroproduction data, let us
apply the $z$ expansion to extract $m_A$ from the inferred $F_A(q^2)$
values for an illustrative dataset, taken from 
Refs.~\cite{Amaldi:1972vf,Brauel:1973cw,Del Guerra:1975wk,DelGuerra:1976uj,Esaulov:1978ed}.    
We have selected datasets that appear in the compilation~\cite{Bernard:2001rs} (cf. Figure~1 of that reference), 
and that also explicitly list inferred values of $F_A(q^2)$ (see also 
\cite{Amaldi:1970tg,Bloom:1973fn,Joos:1976ng,Choi:1993vt,Liesenfeld:1999mv}).   
Figure~\ref{fig:piphoto} displays extractions of $m_A$ in both the $z$ expansion 
and the dipole ansatz (\ref{eq:dipole}) for each of the five datasets.%
\footnote{
For definiteness, where necessary we have chosen one amongst different models
for applied hard-pion corrections: the BNR prescription~\cite{Benfatto:1973pa} in 
\cite{Del Guerra:1975wk,DelGuerra:1976uj,Esaulov:1978ed},   
and the BNR prescription with first form factor assumption in \cite{Amaldi:1972vf}
(``$F_\pi = F_1^V$'' in Table~2 of \cite{Amaldi:1972vf} ).  
We have combined the low-$Q^2$ and high-$Q^2$ data from \cite{Del Guerra:1975wk} 
and \cite{DelGuerra:1976uj} to obtain the Daresbury(1975/1976) data point in Fig.~\ref{fig:piphoto}.    
}
For the larger bound $|a_k|\le 10$,  
the slope of $F_A(q^2)$ is not constrained to be positive 
by each individual dataset,  and we display only the result for $|a_k|\le 5$.   
Applying the $z$ expansion to the entire (17 point) dataset, we find  
\be\label{eq:mApi}
m_A = 0.92^{+ 0.12}_{-0.13} \pm 0.08 \,{\rm GeV} \qquad \mbox{(electroproduction)} \,,
\ee
where the errors are experimental, and from residual shape uncertainty, as in (\ref{eq:MAz}). 
In contrast, a fit of the same data to the dipole ansatz yields $m_A^{\rm dipole}=1.00 \pm 0.02\,{\rm GeV}$. 
These averages are also displayed in the figure. 
We emphasize that our chosen dataset is not exhaustive
We have not attempted to address questions such as correlations 
between different datasets, or uncertainties from model-dependent 
hard-pion corrections.  
We leave a more detailed treatment to future work. 

\section{Summary \label{sec:summary} }

We have presented a model independent description of the axial-vector form 
factor of the nucleon.   
This form factor plays a crucial role 
in neutrino quasielastic scattering at accelerator energies, which
is a basic signal process for neutrino oscillation studies, and is an important 
ingredient in normalizing the neutrino flux at detector locations. 
Recent tensions between measurements in neutrino scattering at different energies, 
and between neutrino scattering and pion electroproduction measurements indicate
a problem in our understanding of this elementary process.  

Several studies have tried to address these discrepancies. 
Modified nuclear models \cite{Butkevich:2010cr, Benhar:2010nx, 
Juszczak:2010ve} 
have been used to find an axial mass close to the MiniBooNE result.  
Other nuclear models include effects of multi-nucleon emission  
\cite{Martini:2009uj, Martini:2010ex, Amaro:2010sd, Amaro:2011qb, 
Nieves:2011pp, Nieves:2011yp},   
and have been reported to obtain better agreement with the differential MiniBooNE 
data from \cite{AguilarArevalo:2010zc}. 
One of these studies~\cite{Nieves:2011yp} reports a dipole axial mass extracted 
from MiniBooNE data in 
agreement with world averages from \cite{Bernard:2001rs,Lyubushkin:2008pe}. 
Another group~\cite{Bodek:2011ps}, modifies the magnetic form 
factor $G_M$ for nucleons bound in carbon but does not change the form factors $G_E$ or $F_A$.
The assumption of the dipole ansatz (\ref{eq:dipole}) is a crucial 
element in many of these studies.%
\footnote{
A parameterization that modifies the dipole behavior at large $Q^2$ 
is presented in \cite{Bodek:2007ym}. 
}  
Our analysis shows that this ansatz introduces
a strong bias in measurements, which must be addressed in order to 
disentangle nucleon-level interactions from nuclear effects.  

Under the assumption of a definite nuclear model (the RFG model, 
summarized in Appendix~\ref{sec:appendix}, with 
parameter values as in Table~\ref{tab:inputs}), 
we extract $m_A$ as defined model-independently in (\ref{eq:mAdef}) from the differential 
MiniBooNE data~\cite{AguilarArevalo:2010zc}.   
The result is displayed in (\ref{eq:MAz}), $m_A = 0.85^{+0.22}_{-0.07} \pm 0.09\,{\rm GeV}$.  
This result may be contrasted with a fit to an illustrative dataset for 
pion electroproduction displayed in (\ref{eq:mApi}), $m_A = 0.92^{+0.12}_{-0.13}\pm 0.08 \,{\rm GeV}$. 
These values may be compared to fits using the dipole ansatz (\ref{eq:dipole}): 
$m_A^{\rm dipole}=1.29\pm 0.05 \,{\rm GeV}$ (neutrino scattering) and 
$m_A^{\rm dipole}=1.00\pm 0.02 \,{\rm GeV}$ (electroproduction). 
A discrepancy is apparent in the dipole ansatz (\ref{eq:dipole}), but
can be ascribed to the unjustified and restrictive assumption on the form factor shape. 
After gaining firm control over the nucleon-level amplitude, nuclear effects can be 
robustly isolated. 
For example, in the context of the RFG model, we extract the result (\ref{epsilonb}) 
for the binding energy parameter $\epsilon_b$. 

The axial mass parameter, or equivalently, the axial radius (\ref{eq:rAdef}), 
is a fundamental parameter of nucleon structure.
The results (\ref{eq:MAz}),(\ref{eq:mApi}) can be expressed as 
\be
r_A = \bigg\{ 
\begin{array}{l}
0.80^{+0.07}_{-0.17} \pm 0.12 \, {\rm fm} \qquad \mbox{(neutrino scattering)} 
\\ 
0.74^{+0.12}_{-0.09} \pm 0.05 \, {\rm fm}  \qquad \mbox{(electroproduction)}  
\end{array} \,.
\ee
More precise measurements in both neutrino scattering and pion electroproduction 
are necessary to substantially reduce the errors on $m_A$, or equivalently $r_A$.  
This would be necessary to provide a model-independent confirmation of 
the convergence of chiral perturbation theory 
corrections based on comparison of electroproduction and neutrino scattering data. 

A related study of the nucleon vector form factors was presented 
in \cite{Hill:2010yb}.  As described there, different expansion ``schemes'' 
are possible.  For example, we may replace 
(\ref{eq:zexpansion}) with  $\phi(t) F_A(t) = \sum_k a_k z(t)^k$, where 
$\phi$ is analytic below $t_{\rm cut}$.   
A choice such as $\phi \sim (1-t/m^{\prime 2})^n$ with $m^\prime \sim {\rm GeV}$ 
could be used to enforce a $1/Q^{2n}$ falloff 
for asymptotic $Q^2$, while retaining the known analytic structure of the form factor.  
Such modifications do not significantly impact the extraction of $m_A$, 
and we have focused on the simplest choice ($t_0=0$ and $\phi=1$).    

Our study indicates that the error on the axial mass parameter 
extracted using the dipole ansatz is underestimated.
While the errors from a model-independent analysis
may be larger, it is essential to study model-independent numbers in order
to draw firm conclusions.  
The simulation of more complicated neutrino scattering processes (e.g. pion and photon production), 
is indirectly affected by enforcing agreement with the quasielastic data. 
It is important for current and future neutrino 
experiments~\cite{Drakoulakos:2004gn,Ayres:2004js,AguilarArevalo:2006se,
Chen:2007zz,Lyubushkin:2008pe,Sanchez:2009zzf,AguilarArevalo:2010zc,Holmes:2010zza,
Abe:2011ks,Choubey:2011zz} 
to converge on consistent values for fundamental neutrino cross sections.  

The analysis presented here can be applied to other neutrino scattering datasets,
involving different nuclear targets, 
and including neutral current scattering and antineutrino scattering.   
It is interesting to extend the analysis of electroproduction 
data; more precise low-energy electroproduction measurements 
have potential to impact the interpretation of future neutrino measurements.
It is also of interest to incorporate model-independent constraints into more 
sophisticated nuclear models. 

%\newpage

\vskip 0.2in
\noindent
{\bf Acknowledgements}
\vskip 0.1in
\noindent
We thank T.~Katori for useful discussions with respect to the MiniBooNE analysis, 
and V.~Bernard, L.~Elouadrhiri and U.-G. Meissner for supplying data corresponding 
to Figure 2 of \cite{Bernard:2001rs}. 
Work supported by NSF Grant 0855039 and DOE grant DE-FG02-90ER40560.

\begin{appendix}

\section{Appendix: RFG model for quasielastic neutrino scattering \label{sec:appendix} }

A number of notations and conventions for the form factors and RFG nuclear model~\cite{Smith:1972xh} 
exist in the literature.  For completeness we collect here the relevant formulas used in our analysis. 

\subsection{Nucleon matrix element of the weak current}

The relevant part of the weak-interaction 
Lagrangian is 
\be 
{\cal L}=\frac{G_F}{\sqrt{2}}V_{ud}\,\bar
{\ell}\gamma^\alpha(1-\gamma_5)\nu\,\bar{u}\gamma_\alpha(1-\gamma_5)d  + {\rm H.c.} \,. 
\ee
The cross section for $\nu(k) + n(p) \to \ell^-(k^\prime)+ p(p^\prime)$ on a free neutron is 
\begin{eqnarray}\label{sigmafree}
\sigma_{\rm free}=\frac1{4|k\cdot p|}\int\frac{d^3{k^\prime}}{(2\pi)^32E_{{\bm k}^\prime}}
\int\frac{d^3{p^\prime}}{(2\pi)^32E_{{\bm p}^\prime}}\overline{\left|{\cal M}^2\right|}(2\pi)^4\delta^4(k+p-k^\prime-p^\prime),
\end{eqnarray}
where the spin-averaged, squared amplitude is 
\begin{eqnarray}
\label{amp2}
\overline{\left|{\cal M}^2\right|}&=&\frac{G_F^2 |V_{ud}|^2 }{4}
L^{\mu\nu} 
\sum_{\rm
  spins}\,\langle p(p^\prime)|\bar{u}\gamma_\mu(1-\gamma_5) d|n(p)\rangle
\langle p(p^\prime)|\bar{u}\gamma_\nu(1-\gamma_5) d|n(p)\rangle^*.
\end{eqnarray}
The leptonic tensor neglecting the neutrino mass is ($\epsilon^{0123}=-1$)
\begin{align}\label{eq:leptonic}
L^{\mu\nu} 
%&= {\rm Tr}[ ( \slash{k}^\prime \gamma^\mu (1-\gamma_5)\slash{k}\gamma^\nu(1-\gamma_5) ] 
= 
8(k^\mu k^{\prime\nu}+k^\nu k^{\prime\mu}-g^{\mu\nu}k\cdot k^\prime 
- i\epsilon^{\mu\nu\rho\sigma}k_\rho k^{\prime}_{\sigma}) \,.
\end{align} 
The hadronic matrix element appearing in (\ref{amp2}) is parameterized by
\begin{align}
\label{FFdef}
\langle p(p^\prime)|\bar{u} \gamma_\mu(1-\gamma_5) d|n(p)\rangle&=
\bar u^{(p)}(p^\prime) \Gamma_\mu(q) u^{(n)}(p) \,,
\end{align}
where $q=k-k^\prime=p^\prime-p$ and we have defined the vertex function
\begin{multline}
\Gamma_\mu(q) 
= \gamma_\mu F_1(q^2) + {i \over 2m_N}\sigma_{\mu\nu}q^\nu F_2(q^2)+\frac{q_\mu}{m_N}F_S(q^2)
+ \gamma_{\mu}\gamma_5 F_A(q^2) + \frac{p_\mu+p^\prime_\mu}{m_N}\gamma_5F_T(q^2)
\\
+{q_\mu \over m_N} \gamma_5 F_P(q^2)  \,.
\end{multline}
We may write the cross section of (\ref{sigmafree}) as  
\be\label{xs}
\sigma_{\rm free} = {G_F^2 |V_{ud}|^2 \over 16 |k\cdot p|} \int{d^3{k^\prime}\over (2\pi)^3 2E_{{\bm k}^\prime}} L^{\mu\nu} \hat{W}_{\mu\nu} \,,
\ee
where the nucleon structure function is 
\be
\hat{W}_{\mu\nu} =\int{d^3{p^\prime}\over (2\pi)^32 E_{{\bm p}^\prime}} (2\pi)^4\delta^4(p-p^\prime+q)  H_{\mu\nu} \,.
\ee
The hadronic tensor is
\be
H_{\mu\nu} = 
{\rm Tr}[  ( \slash{p}^\prime + m_p )\Gamma_\mu(q) (\slash{p}+m_n )\bar{\Gamma}_\nu(q) ] \,,
\ee
where as usual, $\bar{\Gamma} = \gamma^0 \Gamma^\dagger \gamma^0$.  
We may similarly analyze antineutrino scattering,
$\bar{\nu}(k) + p(p) \to \ell^+(k^\prime) + n(p^\prime)$, 
using (\ref{xs}), taking $L^{\mu\nu} \to L^{\nu\mu}$, 
and making the replacements $m_n \leftrightarrow m_p$, $\Gamma_\mu(q) \to \bar{\Gamma}_\mu(-q)$ in $H_{\mu\nu}$.   

Imposing time-reversal invariance shows that $F_i(q^2)$ are real.    
We will assume isospin symmetry in the following, in which case 
$F_S$ and $F_T$ vanish, $m_n=m_p= m_N$, and $\bar{\Gamma}_\mu(-q)=\Gamma_\mu(q)$.  
The hadronic tensor has the time-reversal invariant decomposition 
\be
\label{form1}
H_{\mu\nu} = -g_{\mu\nu} H_1 + {p_\mu p_\nu\over m_N^2} H_2 - i{ \epsilon_{\mu\nu\rho\sigma} \over 2 m_N^2} p^\rho q^\sigma
H_3 + {q_\mu q_\nu \over m_N^2} H_4  
+ {(p_\mu q_\nu + q_\mu p_\nu)\over 2m_N^2 } H_5
\,. 
\ee
The $H_i$'s are expressed in terms of the form factors $F_i$ as 
\begin{eqnarray}
H_1 &=& 8 m_N^2 F_A^2 -2q^2 \left[ (F_1 + F_2)^2 + F_A^2 \right]   
\,,\nl
H_2 &=& H_5 =  8 m_N^2 \left( F_1^2 + F_A^2\right) -2q^2 F_2^2   
\,,\nl
H_3 &=& -16m_N^2\, F_A (F_1 + F_2 ) 
\,,\nl
H_4 &=& -\frac{q^2}{2}\left(F_2^2 + 4F_P^2 \right)-2m_N^2 F_2^2 
- 4m_N^2 \left( F_1F_2 + 2 F_A F_P \right) 
\,.
\end{eqnarray}
Expressions for complex $F_i$ and nonzero $F_S$,$F_T$ can be found, for example, in \cite{Kuzmin:2007kr}. 

\subsection{Model for the nuclear matrix element}

We employ a standard treatment of nuclear effects,  
the ``Relativistic Fermi Gas" (RFG) model as
presented by Smith and Moniz in \cite{Smith:1972xh}, based on the model presented in
\cite{Moniz:1969sr}. 

We assume that there are $A$ nucleons inside the nucleus, with $A/2$
neutrons and $A/2$ protons. The incoming neutrino interacts with a
neutron with 3-momentum ${\bm p}$, determined by some distribution $n_i({\bm p})$. 
The final state proton phase space is 
limited by a factor of $[1-n_f({\bm p^\prime})]$ enforcing Fermi statistics.
Symbolically,
\be 
\sigma_{\rm nuclear}=n_i({\bm p})\otimes\sigma_{\rm free}(\bm{p} \to \bm{p}^\prime)
\otimes [1-n_f({\bm p^\prime})], 
\ee 
and more explicitly
\begin{align}\label{firstpass}
\sigma_{\rm nuclear} 
&\approx 2V \int{d^3{p} \over (2\pi)^3} n_i(\bm{p}) 
\nl
&\bigg\{ 
{G_F^2\over 16 |k\cdot p|} \int{d^3 {k}^\prime \over (2\pi)^3 2 E_{\bm{k}^\prime}}
\int {d^3{p}^\prime \over (2\pi)^3 2E_{\bm{p}^\prime}} (2\pi)^4 
\delta^4( p-p^\prime + q) L^{\mu\nu} H_{\mu\nu} 
\bigg\} 
[1 - n_f(\bm{p}^\prime) ] \,. 
\end{align} 
To arrive at the final model, two modifications are made. 
First, we make the replacement $k\cdot p \to E_{\bm{k}} E_{\bm{p}}$ 
in the prefactor of (\ref{firstpass}).  This replacement ignores a correction
from the nonzero velocity of the initial state nucleon.   It corresponds 
to the model of \cite{Smith:1972xh}, adopted 
by \cite{AguilarArevalo:2010zc}; for definiteness we have followed this convention. 
Second, we incorporate a ``binding energy'', $\epsilon_b$, 
by expressing $H_{\mu\nu}$ as a function of Lorentz 4-vectors 
$p_\mu$, $q_\mu$ as in (\ref{form1})
and then making in (\ref{firstpass}) the replacements
\be
p^0 \to \epsilon_{\bm{p}} \equiv E_{\bm p} - \epsilon_b \,,
\quad
p^{\prime 0} \to \epsilon^\prime_{\bm{p}^\prime}  \equiv E_{\bm{p}^\prime} \,, 
\ee
with $E_{\bm{p}} \equiv \sqrt{m_N^2 + |\bm{p}|^2}$. 
Again, there is some arbitrariness to the insertion 
of $\epsilon_b$ into the formalism; for definiteness we have followed the conventions of \cite{Smith:1972xh}. 
The cross section is then 
\be\label{xsnucleus}
\sigma_{\rm nuclear} = {G_F^2\over 16 |k\cdot p_T|} \int{d^3{k^\prime}\over (2\pi)^3 2E_{{\bm k}^\prime}} L^{\mu\nu} W_{\mu\nu} \,,
\ee
where $p_T^\mu$ is the 4-momentum of the target nucleus with mass $m_T \equiv  A m_N (1-\epsilon_b)$. 
We work in the target rest frame where $p_T^\mu = m_T \delta^\mu_0$.   
The model nuclear structure function $W_{\mu\nu}$ is defined as
\begin{eqnarray}
W_{\mu\nu} \equiv&\int d^3{p}\,f({\bm p},q^0,\bm{q})
H_{\mu\nu}(\epsilon_{\bm{p}}, \bm{p}; q^0, \bm{q} )  \,,
\end{eqnarray}
with 
\be
f({\bm p},q^0,\bm{q})
=
{m_T V\over 4 \pi^2}
 \, n_i(\bm{p}) 
[1-n_f(\bm{p}+\bm{q})] 
{\delta( \epsilon_{\bm{p}} - \epsilon^\prime_{\bm{p}+\bm{q}} + q^0 )  \over \epsilon_{\bm{p}} \epsilon^\prime_{\bm{p}+\bm{q}}} \,. 
\ee
The distribution of neutrons and protons is 
\be\label{niRFG}
n_i(\bm{p}) = \theta( p_F - |\bm{p}| ) \,,
\quad 
n_f(\bm{p}^\prime) = \theta( p_F - |\bm{p}^\prime|) \,,
\ee
where $p_F$ is a parameter of the model.  
The normalization $V$ is fixed by requiring $A/2$ neutrons 
below the Fermi surface (accounting for 2 fermionic spin states), 
\be\label{ninor}
{A\over 2}  = 2V \int{d^3 p \over (2\pi)^3} n_i(\bm{p})  \implies V = {3\pi^2 A\over 2 p_F^3} \,.
\ee
We can expand $W_{\mu\nu}$ in a similar way to $H_{\mu\nu}$ in 
(\ref{form1}):
\be\label{form2}
W_{\mu\nu} = -g_{\mu\nu} W_1 + {p^T_\mu p^T_\nu\over m_T^2} W_2 - i{ \epsilon_{\mu\nu\rho\sigma} \over 2 m_T^2} p_T^\rho q^\sigma
W_3 + {q_\mu q_\nu \over m_T^2} W_4  
+ {(p^T_\mu q_\nu + q_\mu p^T_\nu)\over 2m_T^2 } W_5 \,. 
\ee
The functions $W_i$ are related to integrals over $H_i$.  The
relations can be expressed in terms of the following integrals \cite{Smith:1972xh}: 
\begin{align}
a_1 &= \int d^3 {p} \, f({\bm p},q)\,, 
& a_2 = \int d^3 {p} \, f({\bm p},q)\,{|\bm{p}|^2\over m_N^2}\,,\nl
a_3 &= \int d^3 {p} \, f({\bm p},q)\, {(p^z)^2 \over m_N^2} \,, 
& a_4 = \int d^3 {p} \, f({\bm p},q)\, {\epsilon_{\bm{p}}^2 \over m_N^2} \,,\nl
a_5 &= \int d^3 {p} \, f({\bm p},q)\, {\epsilon_{\bm{p}} p^z \over m_N^2} \,, 
&a_6 = \int d^3 {p} \, f({\bm p},q)\, {p^z \over m_N}\,,\nl
a_7 &=\int d^3 {p} \, f({\bm p},q)\, {\epsilon_{\bm{p}} \over m_N} \,,
\end{align} 
where $|\bm{p}|^2=(p^x)^2+(p^y)^2+(p^z)^2$ and  
the $z$ axis is parallel to $\bm q$.
A straightforward but tedious comparison shows that
\begin{align} \label{Wi}
W_1 &= a_1 H_1 + \frac12( a_2 - a_3) H_2 \,,
\nl
W_2 &= \left[ a_4 + {\omega^2\over |\bm{q}|^2}a_3- 2{\omega\over |\bm{q}|} a_5 + \frac12\left(1-{\omega^2\over |\bm{q}|^2}\right) (a_2 - a_3) \right] H_2 \,,
\nl
W_3 &= {m_T \over m_N} \left( a_7 - {\omega \over |\bm{q}|} a_6 \right) H_3 \,,
\nl
W_4 &= {m_T^2\over m_N^2} \left[ a_1 H_4 + {m_N\over |\bm{q}|} a_6 H_5 + {m_N^2\over 2 |\bm{q}|^2} (3a_3 - a_2) H_2 \right] \,,
\nl
W_5 &= {m_T\over m_N} \left( a_7 - {\omega\over |\bm{q}|} a_6 \right) H_5 + {m_T\over |\bm{q}|}\left[ 2 a_5 + {\omega \over |\bm{q}|}(a_2 - 3 a_3) \right] H_2 \,,
\end{align}
where we are using $\omega = q^0$.  Recall that the $H_i$ are functions of $q^2=\omega^2-|\bm{q}|^2$.  
For the integrals $a_i$ let us define $\omega_{\rm eff} = \omega - \epsilon_b$, 
and observe that 
\begin{multline}
\delta(\epsilon_{\bm{p}}-\epsilon_{\bm{p}+\bm{q}} + q^0) 
= \delta(E_{\bm{p}}-E_{\bm{p}+\bm{q}} + \omega_{\rm eff}) 
= {E_{\bm{p}+\bm{q}}\over |\bm{p}||\bm{q}|}\, \delta\bigg( 
\cos\theta_{ \bm{p} \bm{q} } - 
{ \omega_{\rm eff}^2-|\bm{q}|^2 + 2\omega_{\rm eff} E_{\bm{p}}  
\over   2|\bm{p}||\bm{q}|} 
\bigg).
\end{multline}
The integrals $a_i$ can be expressed in terms of  
\begin{align}
b_j = \frac{m_TV}{2\pi |\bm{q}|} \int dE_{\bm{p}} { E_{\bm{p}} \over E_{\bm{p}} - \epsilon_b} \left(E_{\bm{p}}\over m_N\right)^j \,, 
\end{align}
for $j=0,1,2$.  
In particular, 
\begin{align}
b_0 &= \frac{m_T V}{2\pi |\bm{q}|}\left( E + \epsilon_b \log(E-\epsilon_b) \right) \bigg|_{E_{\rm lo}}^{E_{\rm hi}} \,,
\nl
b_1 &= \frac{m_TV}{2\pi m_N|\bm{q}|}\left[\frac12 E^2 + \epsilon_b\left( E + \epsilon_b \log(E-\epsilon_b) \right) \right] \bigg|_{E_{\rm lo}}^{E_{\rm hi}} \,,
\nl
b_2 &= \frac{m_TV}{2\pi m^2_N|\bm{q}|}\left\{ \frac13 E^3 + 
\epsilon_b 
\left[\frac12 E^2 + \epsilon_b\left( E + \epsilon_b \log(E-\epsilon_b) \right) \right]
\right\}
 \bigg|_{E_{\rm lo}}^{E_{\rm hi}} \,.
\end{align}
Up to an overall constant these are the $b_i$'s of \cite{Smith:1972xh}. Introducing 
$c= -{\omega_{\rm eff}/ |\bm{q}|}$, 
$d= -(\omega_{\rm eff}^2 - |\bm{q}|^2)/(2|\bm{q}|m_N)$, 
we can express the $a_i$'s as
\begin{align}
&a_1=b_0\,,\quad 
a_2=b_2-b_0\,,\quad
a_3=c^2b_2+2cdb_1+d^2b_0 \,,\quad
a_4=b_2-\frac{2\epsilon_b}{m_N}b_1+\frac{\epsilon^2_b}{m^2_N}b_0\,,\nl
&a_5=-cb_2+\left(\frac{\epsilon_b}{m_N}c-d\right)b_1+\frac{\epsilon_b}{m_N}db_0
\,,\quad
a_6=-cb_1-db_0\,,\quad a_7=b_1-\frac{\epsilon_b}{m_N}b_0 \,.
\end{align}
The range of integration is restricted by the conditions, 
\begin{align}
E_{\bm{p}} &\le E_F \equiv \sqrt{m_N^2 + p_{F}^2} 
\le E_{\bm{p}+\bm{q}} = E_{\bm{p}}+\omega_{\rm eff} \,, \quad
-1 \le { \omega_{\rm eff}^2 -|\bm{q}|^2 + 2\omega_{\rm eff} E_{\bm{p}}  \over 
 2 |\bm{q}|\sqrt{E_{\bm{p}}^2-m_N^2} } \le 1 
\,.
\end{align} 
The latter condition can be expressed as 
\be
\left({E_{\bm{p}} \over m_N} - {cd+\sqrt{1-c^2+d^2}\over 1-c^2} \right)
\left({E_{\bm{p}} \over m_N} - {cd-\sqrt{1-c^2+d^2}\over 1-c^2} \right) \ge 0 \,.
\ee
Define 
\be
E_{\rm lo} = {\rm max}\left( E_F-\omega_{\rm eff}, \, m_N  {cd+\sqrt{1-c^2+d^2}\over 1-c^2} \right) \,,
\quad
E_{\rm hi} = E_F \,.
\ee
Then if $E_{\rm lo}\ge E_{\rm hi}$, there is no contribution for the given kinematics. 

In the rest frame of the nucleus, let $E_\ell$ and $|\vec{P}_\ell|=\sqrt{E_\ell^2 - m_\ell^2}$ 
be the energy and 3-momentum of the charged lepton, and let 
$\theta_\ell$ be the angle between the $3$-momenta of the leptons.  
From (\ref{xsnucleus}), 
the final expression for the differential cross section of neutrino-nucleus scattering is 
\begin{multline}\label{eq:doublediff}
\frac{d\sigma_{\rm nuclear}}{dE_\ell d\cos\theta_\ell}=
{G_F^2 |\vec{P}_\ell | \over 16 \pi^2 m_T }
\Bigg\{2 (E_\ell -|\vec{P}_\ell |\cos \theta_\ell)\,W_1
+( E_\ell  + |\vec{P}_\ell |\cos \theta_\ell) W_2
\\
\pm \frac{1}{m_T}\Big[(E_\ell -|\vec{P}_\ell |\cos \theta_\ell)(E_\nu+E_\ell )-m^2_\ell \Big]W_3
+ \frac{m_\ell ^2}{m_T^2}(E_\ell -|\vec{P}_\ell |\cos \theta_\ell) W_4- \frac{m_\ell ^2}{m_T}\, W_5\Bigg\}\,,
\end{multline}
where $W_i$ are given in (\ref{Wi}), and 
where the upper (lower) sign is for neutrino (anti-neutrino) scattering.

\end{appendix}


\begin{thebibliography}{99}

\bibitem{Gran:2006jn}
  R.~Gran {\it et al.}  [K2K Collaboration],
  %``Measurement of the quasi-elastic axial vector mass in neutrino-oxygen
  %interactions,''
  Phys.\ Rev.\  D {\bf 74}, 052002 (2006)
  [arXiv:hep-ex/0603034].
  %%CITATION = PHRVA,D74,052002;%%
  
\bibitem{:2007ru}
  A.~A.~Aguilar-Arevalo {\it et al.}  [MiniBooNE Collaboration],
  %``Measurement of muon neutrino quasi-elastic scattering on carbon,''
  Phys.\ Rev.\ Lett.\  {\bf 100}, 032301 (2008)
  [arXiv:0706.0926 [hep-ex]].
  %%CITATION = PRLTA,100,032301;%%
  
\bibitem{AguilarArevalo:2010zc}
  A.~A.~Aguilar-Arevalo {\it et al.} [ MiniBooNE Collaboration ],
  %``First Measurement of the Muon Neutrino Charged Current Quasielastic Double Differential Cross Section,''
  Phys.\ Rev.\  {\bf D81}, 092005 (2010).
  [arXiv:1002.2680 [hep-ex]].

\bibitem{AguilarArevalo:2010cx}
  A.~A.~Aguilar-Arevalo {\it et al.}  [MiniBooNE Collaboration],
  %``Measurement of the Neutrino Neutral-Current Elastic Differential Cross
  %Section on Mineral Oil at $E_\nu \sim 1$ GeV,''
  Phys.\ Rev.\  D {\bf 82}, 092005 (2010)
  [arXiv:1007.4730 [hep-ex]].
  %%CITATION = PHRVA,D82,092005;%%
  
 \bibitem{Lyubushkin:2008pe}
  V.~Lyubushkin {\it et al.}  [NOMAD Collaboration],
  %``A study of quasi-elastic muon neutrino and antineutrino scattering in the
  %NOMAD experiment,''
  Eur.\ Phys.\ J.\  C {\bf 63}, 355 (2009)
  [arXiv:0812.4543 [hep-ex]].
  %%CITATION = EPHJA,C63,355;%% 
  
 \bibitem{Bernard:2001rs}
  V.~Bernard, L.~Elouadrhiri and U.~G.~Meissner,
  %``Axial structure of the nucleon: Topical Review,''
  J.\ Phys.\ G {\bf 28}, R1 (2002)
  [arXiv:hep-ph/0107088].
  %%CITATION = JPHGB,G28,R1;%%
  
\bibitem{Espinal:2007zz}
  X.~Espinal and F.~Sanchez,
  %``Measurement of the axial vector mass in neutrino-carbon interactions at
  %K2K,''
  AIP Conf.\ Proc.\  {\bf 967}, 117 (2007).
  %%CITATION = APCPC,967,117;%%

\bibitem{Dorman:2009zz}
  M.~Dorman  [MINOS Collaboration],
  %``Preliminary results for CCQE scattering with the MINOS near detector,''
  AIP Conf.\ Proc.\  {\bf 1189}, 133 (2009).
  %%CITATION = APCPC,1189,133;%%

\bibitem{Hill:2010yb}
  R.~J.~Hill and G.~Paz,
  %``Model independent extraction of the proton charge radius from electron
  %scattering,''
  Phys.\ Rev.\  D {\bf 82}, 113005 (2010)
  [arXiv:1008.4619 [hep-ph]].
  %%CITATION = PHRVA,D82,113005;%%

\bibitem{Lepage:1980fj}
  G.~P.~Lepage and S.~J.~Brodsky,
  %``Exclusive Processes in Perturbative Quantum Chromodynamics,''
  Phys.\ Rev.\  D {\bf 22}, 2157 (1980).
  %%CITATION = PHRVA,D22,2157;%%

\bibitem{Carlson:1985zu}
  C.~E.~Carlson and J.~L.~Poor,
  %``THE NUCLEON AXIAL VECTOR FORM-FACTOR IN PERTURBATIVE QCD,''
  Phys.\ Rev.\  D {\bf 34}, 1478 (1986).
  %%CITATION = PHRVA,D34,1478;%%
 
 \bibitem{Nakamura:2010}
  K.~Nakamura {\it et al.}  [Particle Data Group],
  %``Review of particle physics,''
  J.\ Phys.\ G {\bf 37}, 075021 (2010).
  
\bibitem{Schwinger}
  J.~Schwinger,
  %``Field theory of unstable particles,''
  Annals Phys.\  {\bf 9}, 169-193 (1960).

  
\bibitem{Hill:2006ub}
  For a review and further references see: 
  R.~J.~Hill,
  %``The Modern description of semileptonic meson form factors,''
{\it In the Proceedings of 4th Flavor Physics and CP Violation Conference (FPCP 2006), Vancouver, British Columbia, Canada, 9-12 Apr 2006, pp
027},
  [arXiv:hep-ph/0606023].
  %%CITATION = ECONF,C060409,027;%%
  
\bibitem{Bourrely:1980gp}
  C.~Bourrely, B.~Machet and E.~de Rafael,
  %``Semileptonic Decays Of Pseudoscalar Particles (M $\to$ M-Prime Lepton
  %Lepton-Neutrino) And Short Distance Behavior Of Quantum Chromodynamics,''
  Nucl.\ Phys.\  B {\bf 189}, 157 (1981).
  %%CITATION = NUPHA,B189,157;%%
%
%\bibitem{Boyd:1994tt}
  C.~G.~Boyd, B.~Grinstein and R.~F.~Lebed,
  % ``Constraints On Form-Factors For Exclusive Semileptonic Heavy To Light Meson
  %Decays,''
  Phys.\ Rev.\ Lett.\  {\bf 74}, 4603 (1995)
  [arXiv:hep-ph/9412324].
  %%CITATION = PRLTA,74,4603;%%
%
%\bibitem{Lellouch:1995yv}
  L.~Lellouch,
 %  ``Lattice-Constrained Unitarity Bounds for $\bar
  %B~0\to\pi~+\ell~-\bar\nu_\ell$ Decays,''
  Nucl.\ Phys.\  B {\bf 479}, 353 (1996)
  [arXiv:hep-ph/9509358].
  %%CITATION = NUPHA,B479,353;%%
%
%\bibitem{Arnesen:2005ez}
  M.~C.~Arnesen, B.~Grinstein, I.~Z.~Rothstein and I.~W.~Stewart,
  % ``A precision model independent determination of |V(ub)| from B --> pi e
  %nu,''
  Phys.\ Rev.\ Lett.\  {\bf 95}, 071802 (2005)
  [arXiv:hep-ph/0504209].
  %%CITATION = PRLTA,95,071802;%%
%
%\bibitem{Boyd:1995sq}
  C.~G.~Boyd, B.~Grinstein and R.~F.~Lebed,
  % ``Model independent determinations of anti-B $\to$ D (lepton), D* (lepton)
  %anti-neutrino form-factors,''
  Nucl.\ Phys.\  B {\bf 461}, 493 (1996)
  [arXiv:hep-ph/9508211].
  %%CITATION = NUPHA,B461,493;%%
%
%\bibitem{Caprini:1997mu}
  I.~Caprini, L.~Lellouch and M.~Neubert,
  % ``Dispersive bounds on the shape of anti-B --> D(*) l anti-nu form
  %factors,''
  Nucl.\ Phys.\  B {\bf 530}, 153 (1998)
  [arXiv:hep-ph/9712417].
  %%CITATION = NUPHA,B530,153;%%
%
%\bibitem{Becher:2005bg}
  T.~Becher and R.~J.~Hill,
  %``Comment on form factor shape and extraction of |V(ub)| from B --> pi l
  %nu,''
  Phys.\ Lett.\  B {\bf 633}, 61 (2006)
  [arXiv:hep-ph/0509090].
  %%CITATION = PHLTA,B633,61;%%
%
%\bibitem{Hill:2006bq}
  R.~J.~Hill,
  %``Constraints on the form factors for K --> pi l nu and implications for
  %|V(us)|,''
  Phys.\ Rev.\  D {\bf 74}, 096006 (2006)
  [arXiv:hep-ph/0607108].
  %%CITATION = PHRVA,D74,096006;%%
%
%\bibitem{Bharucha:2010im}
  A.~Bharucha, T.~Feldmann and M.~Wick,
  %``Theoretical and Phenomenological Constraints on Form Factors for Radiative
  %and Semi-Leptonic B-Meson Decays,''
  arXiv:1004.3249 [hep-ph].
  %%CITATION = ARXIV:1004.3249;%%
%
%\bibitem{Bourrely:2008za}
  C.~Bourrely, I.~Caprini and L.~Lellouch,
  %``Model-independent description of $B\to \pi l\nu$ decays and a determination
  %of $|V_{ub}|$,''
  Phys.\ Rev.\  D {\bf 79}, 013008 (2009)
  [arXiv:0807.2722 [hep-ph]].
  %%CITATION = PHRVA,D79,013008;%%

\bibitem{Smith:1972xh}
  R.~A.~Smith and E.~J.~Moniz,
  %``Neutrino Reactions On Nuclear Targets,''
  Nucl.\ Phys.\  B {\bf 43}, 605 (1972)
  [Erratum-ibid.\  B {\bf 101}, 547 (1975)].
  %%CITATION = NUPHA,B43,605;%%

\bibitem{Moniz:1971mt}
  E.~J.~Moniz, I.~Sick, R.~R.~Whitney, J.~R.~Ficenec, R.~D.~Kephart and W.~P.~Trower,
  %``Nuclear Fermi momenta from quasielastic electron scattering,''
  Phys.\ Rev.\ Lett.\  {\bf 26}, 445 (1971).
  %%CITATION = PRLTA,26,445;%%

\bibitem{Budd:2003wb}
  H.~S.~Budd, A.~Bodek and J.~Arrington,
  %``Modeling quasielastic form-factors for electron and neutrino scattering,''
  arXiv:hep-ex/0308005.
  %%CITATION = HEP-EX/0308005;%%

\bibitem{Nambu:1997wa}
  Y.~Nambu and D.~Lurie,
  %``Chirality conservation and soft pion production,''
  Phys.\ Rev.\  {\bf 125}, 1429 (1962).
  %%CITATION = PHRVA,125,1429;%%
%\bibitem{Nambu:1997wb}
  Y.~Nambu and E.~Shrauner,
  %``Soft pion emission induced by electromagnetic and weak interactions,''
  Phys.\ Rev.\  {\bf 128}, 862 (1962).
  %%CITATION = PHRVA,128,862;%%

\bibitem{Amaldi:1972vf}
  E.~Amaldi {\it et al.},
  %``Axial-vector form-factor of the nucleon from a coincidence experiment on
  %electroproduction at threshold,''
  Phys.\ Lett.\  B {\bf 41}, 216 (1972).
  %%CITATION = PHLTA,B41,216;%%

\bibitem{Brauel:1973cw}
  P.~Brauel {\it et al.},
  %``Pi+ electroproduction on hydrogen near threshold at four-momentum transfers
  %of 0.2, 0.4 and 0.6 gev-squared,''
  Phys.\ Lett.\  B {\bf 45}, 389 (1973).
  %%CITATION = PHLTA,B45,389;%%

\bibitem{Del Guerra:1975wk}
  A.~Del Guerra {\it et al.},
  %``Measurements of Threshold pi+ Electroproduction at Low Momentum Transfer,''
  Nucl.\ Phys.\  B {\bf 99}, 253 (1975).
  %%CITATION = NUPHA,B99,253;%%

\bibitem{DelGuerra:1976uj}
  A.~Del Guerra {\it et al.},
  %``Threshold $\pi^+$ electroproduction at high momentum transfer: a
  %determination of the nucleon axial vector form-factor,''
  Nucl.\ Phys.\  B {\bf 107}, 65 (1976).
  %%CITATION = NUPHA,B107,65;%%

\bibitem{Esaulov:1978ed}
  A.~S.~Esaulov, A.~M.~Pilipenko and Yu.~I.~Titov,
  %``Longitudinal and Transverse Contributions to the Threshold Cross-Section
  %Slope of Single Pion Electroproduction by a Proton,''
  Nucl.\ Phys.\  B {\bf 136}, 511 (1978).
  %%CITATION = NUPHA,B136,511;%%

\bibitem{Amaldi:1970tg}
  E.~Amaldi {\it et al.},
  %``On pion electroproduction at 5 fm-squared near threshold,''
  Nuovo Cim.\  A {\bf 65}, 377 (1970).
  %%CITATION = NUCIA,65A,377;%%

\bibitem{Bloom:1973fn}
  E.~D.~Bloom {\it et al.},
  %``MEASUREMENTS OF INELASTIC ELECTRON SCATTERING CROSS-SECTIONS NEAR ONE PION
  %THRESHOLD,''
  Phys.\ Rev.\ Lett.\  {\bf 30}, 1186 (1973).
  %%CITATION = PRLTA,30,1186;%%

\bibitem{Joos:1976ng}
  P.~Joos {\it et al.},
  %``Determination of the Nucleon Axial Vector Form-Factor from pi Delta
  %Electroproduction Near Threshold,''
  Phys.\ Lett.\  B {\bf 62}, 230 (1976).
  %%CITATION = PHLTA,B62,230;%%

\bibitem{Choi:1993vt}
  S.~Choi {\it et al.},
  %``Axial and pseudoscalar nucleon form-factors from low-energy pion
  %electroproduction,''
  Phys.\ Rev.\ Lett.\  {\bf 71}, 3927 (1993).
  %%CITATION = PRLTA,71,3927;%%

\bibitem{Liesenfeld:1999mv}
  A.~Liesenfeld {\it et al.}  [A1 Collaboration],
  %``A Measurement of the axial form-factor of the nucleon by the p(e, e-prime
  %pi+)n reaction at W = 1125-MeV,''
  Phys.\ Lett.\  B {\bf 468}, 20 (1999)
  [arXiv:nucl-ex/9911003].
  %%CITATION = PHLTA,B468,20;%%

\bibitem{Benfatto:1973pa}
  G.~Benfatto, F.~Nicolo and G.~C.~Rossi,
  %``Implications of gauge invariance for pion electroproduction at threshold,''
  Nucl.\ Phys.\  B {\bf 50}, 205 (1972).
  %%CITATION = NUPHA,B50,205;%%
%\bibitem{Benfatto:1973dq}
  G.~Benfatto, F.~Nicolo and G.~C.~Rossi,
  %``Pion electroproduction at threshold,''
  Nuovo Cim.\  A {\bf 14}, 425 (1973).
  %%CITATION = NUCIA,14A,425;%%

\bibitem{Butkevich:2010cr}
 A.~V.~Butkevich,
 %``Analysis of flux-integrated cross sections for quasi-elastic neutrino 
%charged-current scattering off $^{12}$C at MiniBooNE energies,''
 Phys.\ Rev.\  {\bf C82}, 055501 (2010).
 [arXiv:1006.1595 [nucl-th]].

 \bibitem{Benhar:2010nx}
 O.~Benhar, P.~Coletti, D.~Meloni,
 %``Electroweak nuclear response in quasi-elastic regime,''
 Phys.\ Rev.\ Lett.\  {\bf 105}, 132301 (2010).
 [arXiv:1006.4783 [nucl-th]].

 \bibitem{Juszczak:2010ve}
 C.~Juszczak, J.~T.~Sobczyk, J.~Zmuda,
 %``On extraction of value of axial mass from MiniBooNE neutrino quasi-elastic 
%double differential cross section data,''
 Phys.\ Rev.\  {\bf C82}, 045502 (2010).
 [arXiv:1007.2195 [nucl-th]].

\bibitem{Martini:2009uj}
 M.~Martini, M.~Ericson, G.~Chanfray, J.~Marteau,
 %``A Unified approach for nucleon knock-out, coherent and incoherent pion 
%production in neutrino interactions with nuclei,''
 Phys.\ Rev.\  {\bf C80}, 065501 (2009).
 [arXiv:0910.2622 [nucl-th]].

 \bibitem{Martini:2010ex}
 M.~Martini, M.~Ericson, G.~Chanfray, J.~Marteau,
 %``Neutrino and antineutrino quasielastic interactions with nuclei,''
 Phys.\ Rev.\  {\bf C81}, 045502 (2010).
 [arXiv:1002.4538 [hep-ph]].

 \bibitem{Amaro:2010sd}
 J.~E.~Amaro, M.~B.~Barbaro, J.~A.~Caballero, T.~W.~Donnelly, 
C.~F.~Williamson,
 %``Meson-exchange currents and quasielastic neutrino cross sections in the 
%SuperScaling Approximation model,''
 Phys.\ Lett.\  {\bf B696}, 151-155 (2011).
 [arXiv:1010.1708 [nucl-th]].

\bibitem{Amaro:2011qb}
  J.~E.~Amaro, M.~B.~Barbaro, J.~A.~Caballero, T.~W.~Donnelly, J.~M.~Udias,
  %``Relativistic analyses of quasielastic neutrino cross sections at MiniBooNE kinematics,''
  Phys.\ Rev.\  {\bf D84}, 033004 (2011).
  [arXiv:1104.5446 [nucl-th]].

 \bibitem{Nieves:2011pp}
 J.~Nieves, I.~Ruiz Simo, M.~J.~Vicente Vacas,
 %``Inclusive Charged--Current Neutrino--Nucleus Reactions,''
 Phys.\ Rev.\  {\bf C83}, 045501 (2011).
 [arXiv:1102.2777 [hep-ph]]

 \bibitem{Nieves:2011yp}
 J.~Nieves, I.~R.~Simo, M.~J.~V.~Vacas,
 %``The nucleon axial mass and the MiniBooNE Quasielastic Neutrino-Nucleus 
%Scattering problem,''
 [arXiv:1106.5374 [hep-ph]].

\bibitem{Bodek:2011ps}
  A.~Bodek and H.~Budd,
  %``Neutrino Quasielastic Scattering on Nuclear Targets: Parametrizing
  %Transverse Enhancement (Meson Exchange Currents),''
  Eur.\ Phys.\ J.\  C {\bf 71}, 1726 (2011)
  [arXiv:1106.0340 [hep-ph]].
  %%CITATION = EPHJA,C71,1726;%%

\bibitem{Bodek:2007ym}
  A.~Bodek, S.~Avvakumov, R.~Bradford and H.~S.~Budd,
  %``Vector and Axial Nucleon Form Factors:A Duality Constrained
  %Parameterization,''
  Eur.\ Phys.\ J.\  C {\bf 53}, 349 (2008)
  [arXiv:0708.1946 [hep-ex]].
  %%CITATION = EPHJA,C53,349;%%

\bibitem{Drakoulakos:2004gn}
  D.~Drakoulakos {\it et al.}  [Minerva Collaboration],
  %``Proposal to perform a high-statistics neutrino scattering experiment using
  %a fine-grained detector in the NuMI beam,''
  arXiv:hep-ex/0405002.
  %%CITATION = HEP-EX/0405002;%%
%\bibitem{Harris:2004iq}
  D.~A.~Harris {\it et al.}  [MINERvA Collaboration],
  %``Neutrino scattering uncertainties and their role in long baseline
  %oscillation experiments,''
  arXiv:hep-ex/0410005.
  %%CITATION = HEP-EX/0410005;%%

\bibitem{Ayres:2004js}
  D.~S.~Ayres {\it et al.}  [NOvA Collaboration],
  %``NOvA: Proposal to build a 30 kiloton off-axis detector to study nu(mu) --->
  %nu(e) oscillations in the NuMI beamline,''
  arXiv:hep-ex/0503053.
  %%CITATION = HEP-EX/0503053;%%

\bibitem{AguilarArevalo:2006se}
  A.~A.~Aguilar-Arevalo {\it et al.}  [SciBooNE Collaboration],
  %``Bringing the SciBar detector to the booster neutrino beam,''
  arXiv:hep-ex/0601022.
  %%CITATION = HEP-EX/0601022;%%

\bibitem{Chen:2007zz}
  H.~Chen {\it et al.}  [MicroBooNE Collaboration],
  %``Proposal for a New Experiment Using the Booster and NuMI Neutrino
  %Beamlines: MicroBooNE,''
  %%CITATION = 
  Fermilab proposal 0974.

\bibitem{Sanchez:2009zzf}
  M.~C.~Sanchez  [LBNE DUSEL Collaboration],
  %``A Very Long-Baseline Neutrino Experiment from FNAL to DUSEL,''
  AIP Conf.\ Proc.\  {\bf 1222}, 479 (2010).
  %%CITATION = APCPC,1222,479;%%
%\bibitem{Barger:2007yw}
  V.~Barger {\it et al.},
  %``Report of the US long baseline neutrino experiment study,''
  arXiv:0705.4396 [hep-ph].
  %%CITATION = ARXIV:0705.4396;%%

\bibitem{Holmes:2010zza}
  S.~D.~Holmes  [Project X Collaboration],
  %``Project X: A Multi-MW Proton Source at Fermilab,''
{\it In the Proceedings of 1st International Particle Accelerator Conference: IPAC'10, Kyoto, Japan, 23-28 May 2010, pp TUYRA01}.  See also: https://www.ids-nf.org/wiki/FrontPage/Documentation/IDR . 
  %%CITATION = CONFP,C100523,TUYRA01;%%

\bibitem{Abe:2011ks}
  K.~Abe {\it et al.}  [T2K Collaboration],
  %``The T2K Experiment,''
  arXiv:1106.1238 [Unknown].
  %%CITATION = ARXIV:1106.1238;%%
%\bibitem{Abe:2011sj}
  K.~Abe {\it et al.} [ T2K Collaboration ],
  %``Indication of Electron Neutrino Appearance from an Accelerator-produced Off-axis Muon Neutrino Beam,''
  Phys.\ Rev.\ Lett.\  {\bf 107}, 041801 (2011).
  [arXiv:1106.2822 [hep-ex]].

\bibitem{Choubey:2011zz}
  S.~Choubey {\it et al.},
  ``International Design Study for the Neutrino Factory, Interim Design Report,'' IDS-NF-20, March 2011.
  %%CITATION = IDS-NF-20;%%

\bibitem{Kuzmin:2007kr}
  K.~S.~Kuzmin, V.~V.~Lyubushkin and V.~A.~Naumov,
  %``Quasielastic axial-vector mass from experiments on neutrino-nucleus
  %scattering,''
  Eur.\ Phys.\ J.\  C {\bf 54}, 517 (2008)
  [arXiv:0712.4384 [hep-ph]].
  %%CITATION = EPHJA,C54,517;%%

\bibitem{Moniz:1969sr}
  E.~J.~Moniz,
  %``Pion electroproduction from nuclei,''
  Phys.\ Rev.\  {\bf 184}, 1154 (1969).
  %%CITATION = PHRVA,184,1154;%%

\end{thebibliography}
\end{document}